\documentstyle[prb,aps,preprint,tighten]{revtex}

\begin{document}
\widetext

\draft

\title{Extension of Fixed-Node Monte Carlo for Lattice Fermions}

\author{D.F.B. ten Haaf and J.M.J. van Leeuwen}
\address{Instituut-Lorentz, Leiden University,
 P. O. Box 9506, 2300 RA Leiden, The Netherlands}

\date{Preprint INLO-PUB-16/95, 5 October 1995}

\maketitle

\begin{abstract}
In a previous paper, we have described a method to perform
Fixed-Node Quantum Monte Carlo calculations for lattice fermions.
In this paper, we present an extension of this method, by which
it is possible to find information on the properties of the exact
ground-state wave function. We give some further illustrations
of the FNMC and the extended methods.
\end{abstract}

\pacs{PACS numbers: 71.20.Ad, 75.10.Lp, 71.10.+x}

\section{Introduction}
Recently, we described a method to perform Quantum Monte Carlo (QMC)
calculations for lattice fermions \cite{Bemmel,Haaf}. We introduced
an effective Hamiltonian, by which the sign problem that is usually
encountered in such systems, can be avoided. We proved that it leads
to an upper bound for the ground-state energy. This Fixed-Node Monte
Carlo (FNMC) method for lattice systems is very much inspired by the
similar method, developed by Ceperley {\em et al.\/}
\cite{CeperleyAlder,Anderson},
for continuous systems.

In this paper we want to show that, in analogy to the method for
continuum problems, it is possible to extend the FNMC method for
lattice systems in order to obtain approximate information about the
true ground state of the system. This {\em nodal relaxation\/} method, as
it is called by Ceperley {\em et al.\/} \cite{CeperleyAlder2}, can be
straighforwardly implemented also on lattice systems. We will give a
formalism for this method, using the (theoretical) results of the FNMC
method described in \cite{Bemmel,Haaf}. Also, we will give a few more
examples of applications of these methods, in comparing our results to
those presented by other authors, for a few systems frequently used in
the literature.

In the next section we will recall the basic formulas of the Green
Function Monte Carlo (GFMC) formalism, and give our notation and
interpretation of the concept of importance sampling. After a few
summarizing remarks on the Fixed-Node Monte Carlo method, we will show
in Section~\ref{sec-power} how the results of this method can be used
to calculate properties of the true ground state of the system. In
Section~\ref{sec-applications} we present a few examples of
applications of both methods, and we conclude this paper by a
discussion on the applicability of the methods and the need for
further investigation of possible wave functions.

A much more elaborate explanation and discussion of the FNMC method
and the extended method, information on the implementation of both
methods, and some more examples of applications, have been
given in Refs.\onlinecite{dannythesis} and\onlinecite{hansthesis}.

\section{Basic formulas of Green Function Monte Carlo}
\label{sec-GFMC}
The GFMC method, as we use it, is based on the {\em projection\/}
operator
\begin{equation} \label{projector}
  {\cal F} = 1 - \tau({\cal H} - w) ,
\end{equation}
by which the ground state of the Hamiltonian $\cal H$ is filtered from
a trial state $|\Psi_{\text{T}}\rangle$, which is known or can be
calculated for each configuration $R$ in the configuration space $\{
R\}$. The parameters $\tau$ and $w$ are to be chosen appropriately.
After applying this operator $n$ times, an approximation
\begin{equation}\label{psigfmc}
  |\Psi^{(n)}\rangle = {\cal F}^n|\Psi_{\text{T}}\rangle
\end{equation}
of the ground state is obtained, which closely resembles the ground
state for large $n$. The energy of this state can be formally
calculated as
\begin{eqnarray}
  E^{(n)} & = & \frac {\langle\Psi_{\text{T}}|{\cal
      H}|\Psi^{(n)}\rangle} {\langle\Psi_{\text{T}}|
    \Psi^{(n)}\rangle}\\ & = & \frac {\sum_{{\cal
        R}}\langle\Psi_{\text{T}}|H|R_{n}\rangle \left[\prod_{i=1}^{n}
    \langle R_{i}|F|R_{i-1}\rangle\right] \langle
    R_0|\Psi_{\text{T}}\rangle} {\sum_{{\cal
        R}}\langle\Psi_{\text{T}}| R_{n}\rangle \left[\prod_{i=1}^{n}
    \langle R_{i}|F|R_{i-1}\rangle\right] \langle
    R_0|\Psi_{\text{T}}\rangle} ,
\label{mixedestimate}
\end{eqnarray}
where ${\cal R}=\{R_{0},R_{1},R_{2},...,R_{n}\}$ is a path in the
configuration space. As we will see in the actual implementation of
this calculation, the wave function has to be known in the end point
of each path, and therefore $\langle\Psi_{\text{T}}|$ must be used in
this expression in stead of $\langle\Psi^{(n)}|$. This is called a
{\em mixed estimate\/} of the energy.

The expression~(\ref{mixedestimate}) can usually not be calculated
exactly. In the Monte Carlo scheme, only a limited number of paths
${\cal R}$ is used for the calculation of this expression. We can
rewrite~(\ref{mixedestimate}) as
\begin{equation}
  E^{(n)} = \frac{\sum_{{\cal R}} E_{\text{T}}(R_n)m({\cal R})p({\cal
      R})}{\sum_{{\cal R}} m({\cal R})p({\cal R})} ,
\end{equation}
where
\begin{equation}
  E_{\text{T}}(R) =
  \frac{\langle\Psi_{\text{T}}|H|R\rangle}{\langle\Psi_{\text{T}}|R\rangle}
\end{equation}
is the {\em local energy\/} in the configuration $R$.  The quantity
$p({\cal R})$ can be interpreted as the probability of choosing a
complete path ${\cal R}$, and each path carries a {\em weight\/} or
{\em multiplicity\/} $m({\cal R})$. If one makes sure that $p$ and $m$
satisfy
\begin{equation}\label{pmrelation}
  p({\cal R})m({\cal R}) =
  \langle\Psi_{\text{T}}|R_{n}\rangle\left[\prod_{i=1}^{n} \langle
  R_{i}|F|R_{i-1}\rangle\right]\langle R_0|\Psi_{\text{T}}\rangle ,
\end{equation}
the energy $E^{(n)}$ can be calculated as
\begin{equation}\label{gfmcenergy}
  E^{(n)}_{\text{GFMC}} = \frac {\sum_{{\cal R}}^\prime
    E_{\text{T}}(R_n)m({\cal R})} {\sum_{{\cal R}}^\prime m({\cal R})}
  ,
\end{equation}
where the summation is now restricted to some number of paths, chosen
according to the probability $p({\cal R})$. In the limit of using
infinitely many paths, this calculation becomes exact; in case of a
finite number of paths, a statistical error bar can be defined.

In the definition of $p$ and $m$ one can include ways to make the
sampling of the wave function more efficient. This is called {\em
  importance sampling\/}, and usually refers to the idea that one
should try to sample more often in regions where the trial wave
function has a large absolute value. We give a description which
includes the possibility of performing importance sampling by means of
a guiding function $\Psi_{\text{G}}$. This function must be positive
everywhere; often, it is taken to be the absolute value of the trial
wave function. We denote the probability for choosing a path as
\begin{equation} \label{guidingprobability}
  p_{\text{guiding}}({\cal R}) =
  p_{0}(R_{0})\prod_{i=1}^{n}p(R_{i}\leftarrow R_{i-1}) ,
\end{equation}
Here, the probability for the first configuration of the path to be
$R$ is
\begin{equation} \label{startingprob}
  p_{0}(R) = \langle\Psi_{\text{T}}|R\rangle\langle
  R|\Psi_{\text{T}}\rangle ,
\end{equation}
and the probability of each subsequent step of the path is defined by
\begin{equation} \label{hoppingprob}
  p(R^\prime\leftarrow R) =
  \frac{\langle\Psi_{\text{G}}|R^\prime\rangle\langle
    R^\prime|F|R\rangle} {\langle\Psi_{\text{G}}|R\rangle m(R)} .
\end{equation}
When using this probability for the paths, one has to take $m$ as
\begin{equation} \label{guidingweight}
  m_{\text{guiding}}({\cal R}) =
\frac{\langle\Psi_{\text{G}}|R_{0}\rangle}{\langle\Psi_{\text{T}}|R_{0}\rangle}
  \left[\prod_{i=1}^{n}m(R_{i-1})\right]
\frac{\langle\Psi_{\text{T}}|R_{n}\rangle}{\langle\Psi_{\text{G}}|R_{n}\rangle}
    ,
\end{equation}
where the weight factors $m(R_{i-1})$ can be calculated for each step
$R_i \leftarrow R_{i-1}$ separately as
\begin{equation} \label{hoppingmult}
  m(R) = \sum_{R^\prime} \frac{\langle\Psi_{\text{G}}|R^\prime\rangle
    \langle R^\prime|F|R\rangle}{\langle\Psi_{\text{G}}|R\rangle} =
  \frac{\langle\Psi_{\text{G}}|F|R\rangle}{\langle\Psi_{\text{G}}|R\rangle}
  .
\end{equation}
These relations basically define the Monte Carlo procedure.
Additionally, one can make use of {\em branching\/}
\cite{TrivediCeperley,CeperleyAlder},
to avoid that the variance of the calculated result increases too
rapidly.

In the Monte Carlo scheme, one can thus choose the starting
configurations of the paths according to $p_0$, and each subsequent
configuration according to the {\em stochastic matrix\/} defined by
$p(R^\prime\leftarrow R)$. Interpreting the paths as random walks in
the configuration space, one usually calls the configurations chosen
{\em walkers\/}. In our interpretation, the starting set of walkers
that one uses in the Monte Carlo calculations is distributed according
to the square of the trial wave function. One can calculate how the
walkers are redistributed after $n$ steps, considering also the weight
they carry:
\begin{equation}
  p(R) = \sum_{\{{\cal R}|R_{n}=R\} } m({\cal R})p({\cal R}) .
\end{equation}
This expression can be easily evaluated using the
relation~(\ref{pmrelation}) for $p$ and $m$:
\begin{eqnarray}
  p(R) & = & \langle\Psi_{\text{T}}|R\rangle
  \sum_{\{R_{0},R_{1},\cdots,R_{n-1}\}} \left[\prod_{i=1}^{n} \langle
  R_{i}|F|R_{i-1}\rangle\right]\langle R_0|\Psi_{\text{T}}\rangle\\ &
  = & \langle\Psi_{\text{T}}|R\rangle\langle R|{\cal
    F}^{n}|\Psi_{\text{T}}\rangle \\ & = &
  \langle\Psi_{\text{T}}|R\rangle\langle R|\Psi^{(n)}\rangle .
  \label{walkersdistribution}
\end{eqnarray}
Thus, after a sufficiently large number of steps the walkers are
distributed according to the product of the trial wave function and
the ground-state wave function of the Hamiltonian considered.

\section{Extension of the FNMC formalism}
\label{sec-power}
The Fixed-Node Monte Carlo method uses in fact the same formalism as
the general Green Function Monte Carlo, presented in the previous
section, but with a modified Hamiltonian ${\cal H}_{\text{eff}}$ to
avoid steps that could introduce a change of sign in the weights or
the probabilities (the {\em sign problem\/}\cite{signproblem}). The
GFMC procedure is straightforwardly applied to this {\em effective\/}
Hamiltonian, for which a prescription has been given in
Refs.\onlinecite{Bemmel} and\onlinecite{Haaf}. Thus, by the FNMC method,
information is
obtained not on the ground state of the Hamiltonian ${\cal H}$ one is
interested in, but on the ground state of ${\cal H}_{\text{eff}}$. It
has been proven that the ground-state energy of ${\cal H}_{\text{eff}}$
provides an upper bound for the ground-state energy
of ${\cal H}$. Of course, it would be better to obtain information
directly on ${\cal H}$. As we will see, under certain circumstances
this is possible, using the FNMC result as input for a new
calculation, called {\em nodal relaxation\/}. It strongly resembles
the method described by
Ceperley and Alder \cite{CeperleyAlder} for a continuum problem. The
method suffers from the sign problem, but it may yield convergent
results before the fluctuations destroy the accuracy of the
calculations.

Let us rewrite the expression~(\ref{walkersdistribution}) to give the
distribution of the configurations in the FNMC procedure after $n$
steps:
\begin{equation}
  p_{\text{FN}}^{(n)}(R) = \langle\Psi_{\text{T}}|R\rangle\langle
  R|\Psi_{\text{eff}}^{(n)}\rangle ,
\end{equation}
where $|\Psi_{\text{eff}}^{(n)}\rangle$ is supposed to be a good
approximation of the ground state $|\Psi_{\text{eff}}\rangle$ of ${\cal
H}_{\text{eff}}$. The idea is now that
$|\Psi_{\text{eff}}\rangle$ is sufficiently close to the ground state
$|\Psi_0\rangle$ of ${\cal H}$, such that in GFMC procedure used on
${\cal H}$ convergence can be obtained within a relatively small
number of steps. Thus, we modify~(\ref{psigfmc}) in the following way:
\begin{equation}
  |\Psi_{\text{P}}^{(n)}\rangle = {\cal F}^n|\Psi_{\text{eff}}\rangle
  .
\end{equation}
Then, we also have to modify the expression~(\ref{mixedestimate}) for
the mixed estimate of the energy:
\begin{eqnarray}
  E_{\text{P}}^{(n)} & = & \frac {\langle\Psi_{\text{T}}|{\cal
      H}|\Psi_{\text{P}}^{(n)}\rangle} {\langle\Psi_{\text{T}}|
    \Psi_{\text{P}}^{(n)}\rangle}\\ & = & \frac {\sum_{{\cal
        R}}\langle\Psi_{\text{T}}|H|R_{n}\rangle \left[\prod_{i=1}^{n}
    \langle R_{i}|F|R_{i-1}\rangle\right] \langle
    R_0|\Psi_{\text{eff}}\rangle} {\sum_{{\cal
        R}}\langle\Psi_{\text{T}}| R_{n}\rangle \left[\prod_{i=1}^{n}
    \langle R_{i}|F|R_{i-1}\rangle\right] \langle
    R_0|\Psi_{\text{eff}}\rangle} .
\end{eqnarray}
Note that we still have to put $\langle\Psi_{\text{T}}|$ in this
expression, as that is the only available information we have in each
configuration. However, in the starting configurations of the paths,
the trial wave function has been replaced by the effective
ground-state wave function, of which we have information through the
distribution of the walkers in the FNMC procedure. We have to pay
further attention to the fact that sign changes can now occur, which
have to be embedded in the transition probabilities and
multiplicities. We can use the following guiding procedure, adapted
from the regular GFMC scheme described in the previous section: the
probability to start in a configuration $R$ is given by the fixed-node
resulting distribution:
\begin{equation}\label{powerinputdistribution}
  p_{0}(R) = p_{\text{FN}}^{(n)}(R) .
\end{equation}
The transition probabilities are given by
\begin{equation}
  p(R^\prime\leftarrow R) =
  \frac{\langle\Psi_{\text{G}}|R^\prime\rangle
    \left|\rule{0mm}{2ex}\langle
      R^\prime|F|R\rangle\right|}{\langle\Psi_{\text{G}}|R\rangle
      m(R)} ,
\end{equation}
with the modified weight factor
\begin{equation}
  m(R) = \sum_{R^\prime} \frac{\langle\Psi_{\text{G}}|R^\prime\rangle
    \left|\rule{0mm}{2ex}\langle
      R^\prime|F|R\rangle\right|}{\langle\Psi_{\text{G}}|R\rangle} .
\end{equation}
The absolute value has to be taken to make sure that the transition
probabilities remain positive, such that the procedure can still be
interpreted as a stochastic walk. With these expressions
replacing~(\ref{startingprob}), (\ref{hoppingprob}),
and~(\ref{hoppingmult}), the equations~(\ref{guidingprobability})
and~(\ref{guidingweight}) can again be used for the total probability
and the total ``weight'' of a path, respectively. Note that the
weights can now be negative, due to the possibly different signs of
the wave function in the starting and end configurations of the path.

By this formalism, one can thus try to obtain information directly on
the true ground state of the Hamiltonian, using the fixed-node result
as a starting point. In the following section, we will show a few
examples of applications of this method.

\section{Applications}
\label{sec-applications}
First, we visualize how the FNMC
and the nodal-relaxation method work for the Hubbard model on
a small system, the
2 $\times$ 2 $\times$ 2 cube, with $U=2.5$. We have performed exact
calculations for
this system\cite{cube}, in order to be able to compare the Monte Carlo
results and check whether the program has been correctly implemented.
In Figures~\ref{fig1} and~\ref{fig2}, we show calculations for
this system at half filling and with zero total spin in the $z$-direction.
In Figure~\ref{fig1}, the first 250 steps in the Fixed-Node Monte Carlo
calculation, using a homogeneous mean-field wave function as trial
wave function, are shown. As can be seen, after some 100 steps the
energy measured starts fluctuating around the exact ground-state
energy of the effective Hamiltonian (indicated by the drawn line).
The square at the right indicates the energy that one obtaines after
averaging over several thousands of steps (not shown in this figure),
with a
statistical error which is smaller than the symbol size.
In Figure~\ref{fig2}, two separate calculations of nodal relaxation are shown;
one using the same trial wave function as in Figure~\ref{fig1} (indicated by
H), the other using a mean-field wave function with antiferromagnetic ordering
imposed on the spins (indicated by AF). The horizontal lines indicate the exact
ground-state energies of the effective (FN) Hamiltonians using the H and the AF
trial functions, respectively, and of the true Hamiltonian. The resulting
energy after each
step is indicated by an error bar (the energy measured being the middle value
of the error bar). As can be seen, the size of the error bars in both
calculations first decreases with increasing number of steps, and then starts
increasing, indicating the existence of the sign problem. The calculation
marked H seems to converge to the correct value before the sign problem arises;
the AF calculation does not has not converged at that point. Clearly, this is
due to the fact that the fixed-node energy found when using the AF wave
function is significantly higher than the resulting energy in the other
fixed-node calculation. This nicely demonstrates that the exact result can be
reached, but only if the fixed-node result is close enough to the true ground
state.

We have performed some more calculations for the Hubbard model.
In Table~\ref{tab1}, we compare results of calculations of the ground-state
energy for a 4 $\times$ 4 square with periodic boundary conditions (PBC), with
5 spins up and 5 down, and $U=4$. For this system, exact and approximate
results are available in the literature. As can be seen, in the case of $U=4$,
the fixed-node method yields an upper bound for the ground-state energy which
is quite close to the exact value, and nodal relaxation leads to the exact
result with an error bar of only 0.05\% . In Table~\ref{tab2}, we show the
results of various calculations of the ground-state energy in a 10 $\times$ 10
system (PBC), at half filling and with zero total spin in the $z$-direction,
also for $U=4$. There are no exact results available for this system, but
several results of mean-field and quantum Monte Carlo calculations can be found
in the literature. As can be seen, our results are in very good agreement with
previous QMC results by Hirsch \cite{Hirsch} and White {\em et
al.\/}\cite{White}.

\section{Discussion and outlook}
First we would like to make a few additional remarks on the calculations we
have performed. As one can see from Table~\ref{tab1}, adding a Gutzwiller
factor to the bare mean-field wave function greatly improves the (variational)
energy obtained from the mean-field calculation. In the fixed-node calculation,
however, the energy is hardly improved by adding the Gutzwiller factor to the
trial wave function. Considering also other calculations we performed, we feel
that this is a rather general feature of the fixed-node method. In the
fixed-node method, the energy that is obtained is determined by the {\em sign
structure\/} of the trial wave function, via the ratios of the wave function in
neighboring configurations with different signs. In order to significantly
improve the energy, one must be able to change the sign of the trial wave
function in individual configurations, as that is the most important factor
determining the energy that can be obtained. By a Gutzwiller factor (or,
similarly, a Jastrow factor) one does not change the sign, but only the ratios
of the wave function. Another interesting point is that the Monte Carlo
calculations become less stable for larger $U$. This is due to the fact that
the parameter $\tau$ has to be taken smaller in that case, such that
convergence is slowed down. For the fixed-node method this is not a very big
problem, as one can simply sample for a longer time to obtain a result with
desired accuracy, but in the nodal-relaxation method one can not afford to have
longer paths in the calculations, as the sign problem starts showing up after a
certain number of steps.

Another problem concerns the observables that can be calculated. For operators
that commute with the Hamiltonian, similar formulas as have been given for the
energy in Section~\ref{sec-GFMC} can be used, although (in fixed node) the
results obtained for operators other than the Hamiltonian itself can not
present a bound to the exact result. For operators that do not commute with the
Hamiltonian, however, there is a more serious problem; in that case, the mixed
estimate does not yield the correct value of the observable for the ground
state of the (effective or true, depending on the method) Hamiltonian. Usually,
one uses an {\em extrapolated estimate\/}\cite{extrest} in that case, taking
two times the mixed estimate minus the variational result (i.e., the value of
the observable in the trial state). In practice, this leads to rather good
results; however, one is unable to check how good the approximation is.

Recently, Zhang {\em et al.\/}\cite{Zhang} presented the {\em Constrained Path
Monte Carlo\/} method for lattice fermions, which they claim to be a better
method than fixed node. Like we, they apply some kind of fixed-node principle,
but in a {\em continuous\/} space of Slater determinants, which seems to cause
that the CPMC energy is a better approximation of the true energy than the
fixed-node energy. Also, they claim that they have a better method ({\em
backtracing\/}) to perform calculations for non-commuting operators. It is not
very clear to us how CPMC compares to fixed node combined with nodal
relaxation. It seems to us that our method is more generally applicable, as we
can use any wave function as trial wave function (as long as it can be
calculated relatively easily) and we can also treat systems of frustrated
bosons or spins, while their method restricts the trial wave function to be a
Slater determinant. In any case, it would be very interesting to make a better
comparison of both methods for some different systems.

Finally, we state that the most important ingredient of our method is the trial
wave function. The quality of the trial function, i.e., its sign structure,
determines how good the approximation is that can be obtained. This is not a
trivial problem, for, as Ceperley states, ` \ldots little progress [has been
made] in stating exact conditions that nodes must obey.' \cite{Ceperleycite}
Most studies of wave functions aim at improving the energy of the wave function
itself, while for our purposes it would be necessary to explore its sign
structure, in order the improve the energy that results after a fixed-node
Monte Carlo simulation. We still not have a clear understanding, yet, of the
requirements that a wave function must fulfill to make a good fixed-node trial
wave function. However, we feel that the fixed-node method combined with nodal
relaxation offers great new opportunities to tackle problems of lattice
Hamiltonians.

\subsection*{Acknowledgment}
This work is part of the research program of the Stichting voor Fundamenteel
Onderzoek der Materie (FOM), which is financially supported by the Nederlandse
Organisatie voor Wetenschappelijk Onderzoek (NWO).

\begin{table}[tb!]
\caption[]{\label{tab1}
Comparison of the exact ground-state energy to the energy of mean-field and
Gutzwiller wave functions, and to the energy obtained by MC simulations, for
the 4 $\times$ 4 square (PBC), with 5 spins up and 5 down, for $U=4$.
A mean-field (MF) homogeneous (H) wave function has been used. The Gutzwiller
wave function (GMF) was used with $g=0.6$; its energy was calculated by
variational Monte Carlo.
For the fixed-node (FN) and nodal relaxation (NR) Monte Carlo calculations, a
trial wave function has been used as indicated.  Energies are given per site
and in units of $t$.
Constrained Path Monte Carlo (CPMC) results were taken from
Ref.\protect\onlinecite{Zhang}, exact results from
Ref.\protect\onlinecite{Parola}.}
\vspace*{2ex}
\begin{center}
\begin{tabular}{|c|c|c|c|c|c|c|c|}
\hline
$U$ & MF & GMF & FN/MF & FN/GMF & NR/GMF & CPMC & exact \\
\hline \hline
4 & -1.109 & -1.212 & -1.2186(4) & -1.2201(4) & -1.2234(6) & -1.2238(6) &
-1.2238 \\
\hline
\end{tabular} 
\end{center} \vspace*{-3ex}
\end{table}

\begin{table}[bt!]
\caption[]{\label{tab2}
Various mean-field and quantum Monte Carlo calculations (QMC) of the exact
ground-state energy of a 10 $\times$ 10 square (PBC), at half filling and with
zero total $S^z$, for $U=4$.
Slave Boson MF result has been taken from Ref.\protect\onlinecite{Denteneer},
QMC results from Refs.\protect\onlinecite{Hirsch}
and\protect\onlinecite{White}.}
\vspace*{2ex}
\begin{center}
\begin{tabular}{|l|r@{}l|}
\hline
{}~method & \multicolumn{2}{|c|}{~energy per site~~} \\
\hline
{}~Mean Field (AF) & ~-0 & .797 \\
{}~Slave Boson MF~ & -0 & .839 \\
{}~Gutzwiller VMC & -0 & .842 \\
{}~QMC (Hirsch '85)~ & -0 & .88(3) \\
{}~QMC (White '89)~ & -0 & .864(1) \\
{}~FNMC & -0 & .852(2) \\
{}~Power MC~ & -0 & .860(5)~  \\
\hline
\end{tabular} 
\end{center} \vspace*{-3ex}
\end{table}

\begin{figure}
\caption{\label{fig1} First part of a FNMC simulation, for a 2 $\times$
2 $\times$ 2 cube (see text for details). Each star indicates the average
of the energy samples in 10 Monte Carlo steps. The drawn line indicates
the exact ground-state energy of the fixed-node effective Hamiltonian.
The square at the right indicates the resulting energy, obtained after
averaging over several thousands of steps, with a statistical error
which is smaller than the symbol size.}
\end{figure}

\begin{figure}
\caption{\label{fig2} Nodal relaxation for the system as in Fig.~\ref{fig1}.
Two calculations using different trial wave functions are shown
(see text for details). Error bars indicate the estimated values of
the ground-state energy after each step of the calculation.}
\end{figure}

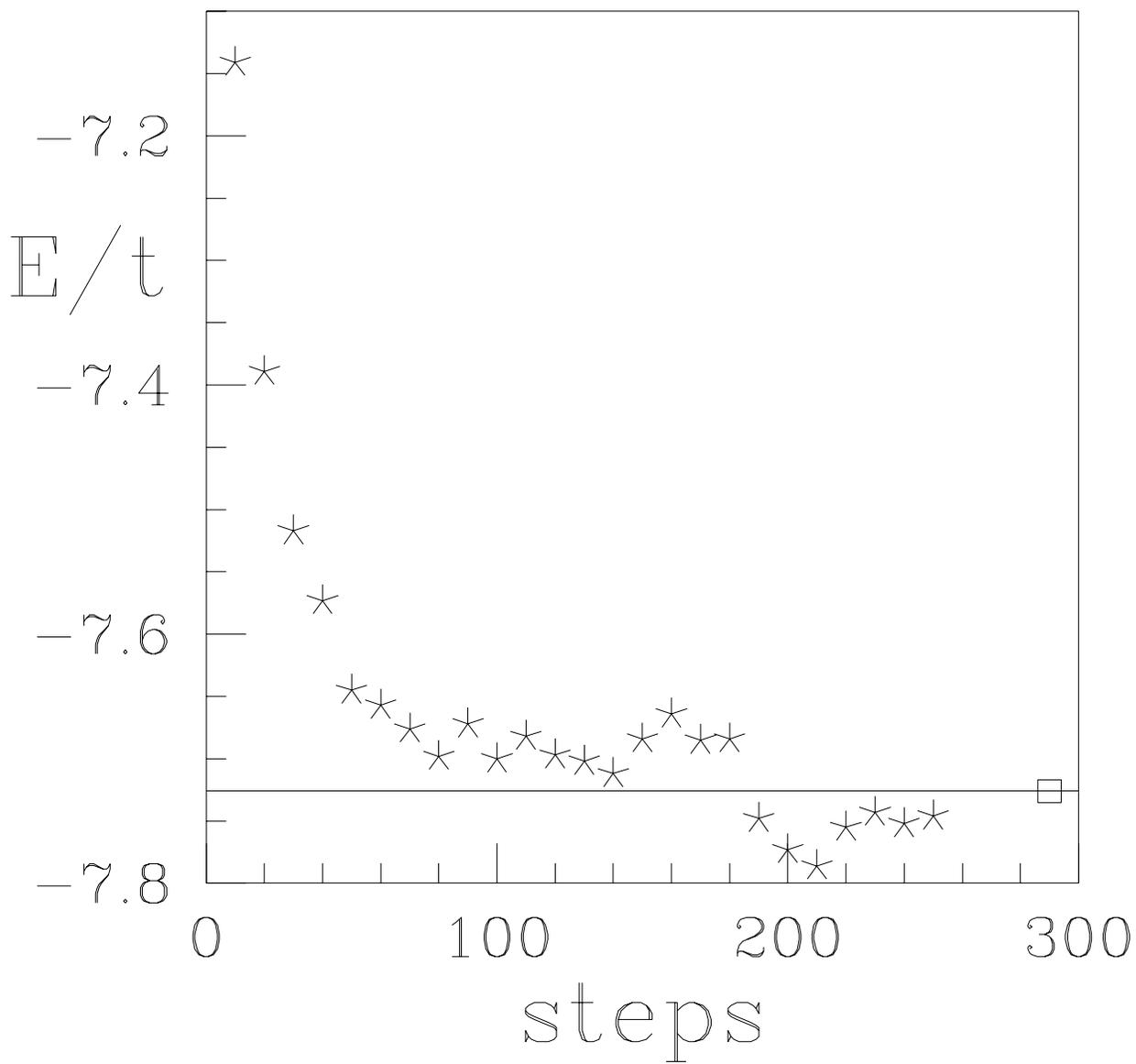
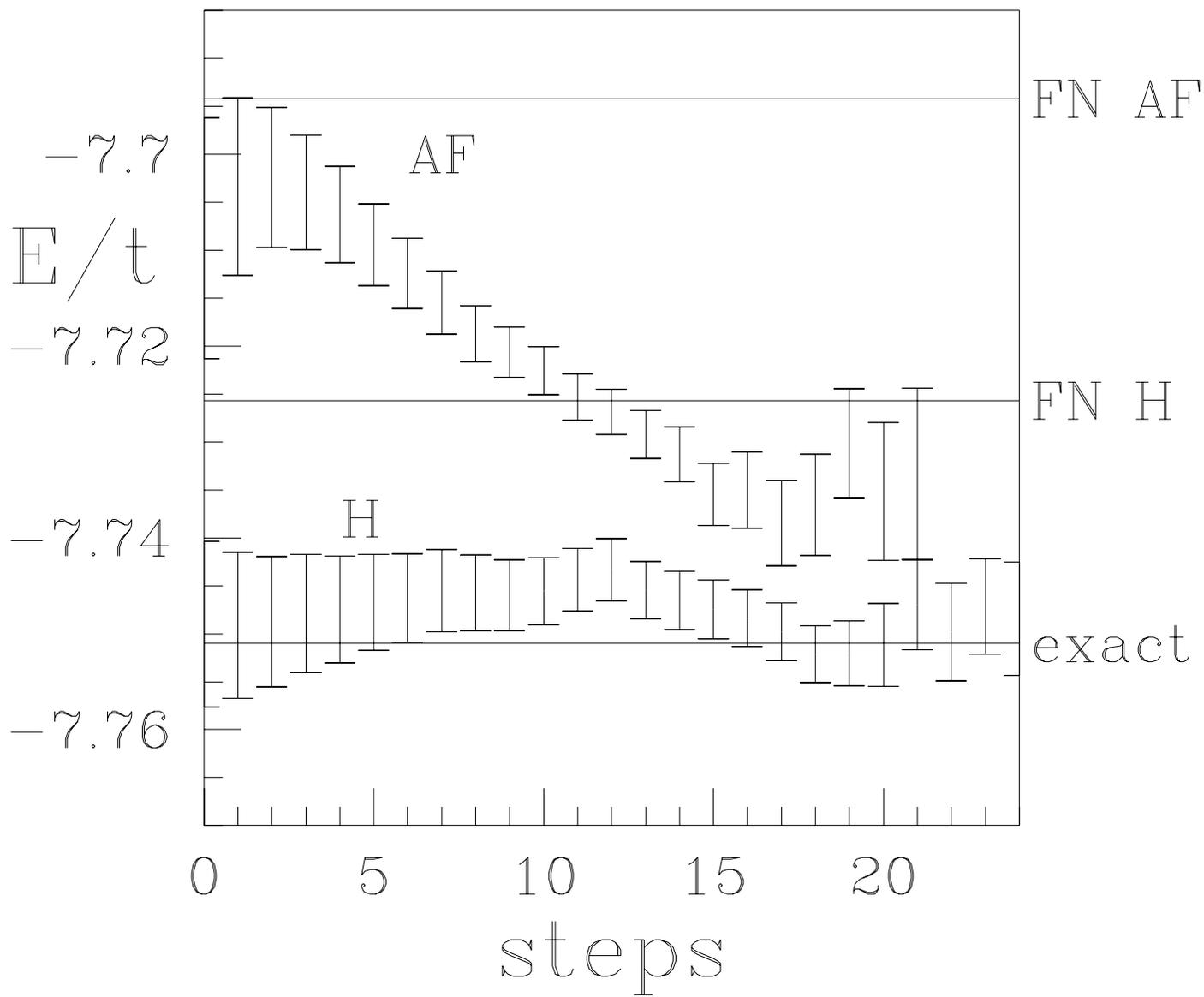
\end{document}